\documentclass[12pt,preprint]{aastex}

\slugcomment{To appear in ApJL}

\shorttitle{Dusty disks at the bottom of the IMF}
\shortauthors{Scholz \& Jayawardhana}

\begin{document}
\bibliographystyle{apj}


\title{Dusty disks at the bottom of the IMF}


\author{Alexander Scholz}
\affil{SUPA, School of Physics \& Astronomy, University of St. Andrews, North Haugh, St. Andrews, 
KY16 9SS, United Kingdom}
\email{as110@st-andrews.ac.uk}
\author{Ray Jayawardhana}
\affil{Department of Astronomy \& Astrophysics, University of Toronto, 50 St. George Street, Toronto, 
ON M5S 3H4, Canada}
\email{rayjay@astro.utoronto.ca}

\begin{abstract}
'Isolated planetary mass objects' (IPMOs) have masses close to or below the Deuterium-burning 
mass limit ($\sim 15\,M_\mathrm{Jup}$) -- at {\it the bottom of the stellar initial mass function}. 
We present an exploratory survey for disks in this mass regime, based on a dedicated 
observing campaign with the Spitzer Space Telescope. Our targets include the full sample of 
spectroscopically confirmed IPMOs in the $\sigma$\,Orionis cluster, a total of 18 sources. In the
mass range $8\ldots 20\,M_\mathrm{Jup}$, we identify 4 objects with $>3\sigma$ colour excess at a 
wavelength of 8.0$\,\mu m$, interpreted as emission from dusty disks. We thus establish that a substantial
fraction of IPMOs harbour disks with lifetimes of at least 2-4\,Myr (the likely age of the cluster),
indicating an origin from core collapse and fragmentation processes. The disk frequency in the IPMO 
sample is $29\pm ^{16}_{13}$\% at 8.0\,$\mu m$, very similar 
to what has been found for stars and brown dwarfs ($\sim 30$\%).
The object SOri\,70, a candidate  $3\,M_\mathrm{Jup}$ object in this cluster, shows IRAC colours 
in excess of the typical values for field T dwarfs (on a 2$\sigma$ level), possibly due to disk 
emission or low gravity. This is a new indication for youth and thus an extremely low mass for 
SOri\,70.

\end{abstract}

\keywords{stars: circumstellar matter, formation, low-mass, brown dwarfs -- planetary systems}

\section{Introduction}
\label{intro}

Over the past five years, it has been firmly established that the evolution of brown 
dwarfs from 1-10\,Myr follows the blueprint known from T Tauri stars: They show evidence
for ongoing gas accretion and outflows \citep[e.g.][]{2005ApJ...625..906M,2005ApJ...626..498M}, and 
harbour circum-sub-stellar disks with lifetimes of 5-10\,Myr
\citep[e.g.][]{2003AJ....126.1515J,2007ApJ...660.1517S}, not vastly different 
from stars. This has often been interpreted as an indication for a common origin of stars and brown 
dwarfs -- sub-stellar objects as the natural extension of the stellar initial mass function (IMF) and
the lowest-mass outcome of core collapse and fragmentation processes
\citep[see review by][and references therein]{2007prpl.conf..443L}. At the same time, ongoing 
deep surveys have pushed the sensitivity limits down to masses below the Deuterium burning 
limit at $\sim 15\,M_{\mathrm{Jup}}$, finding evidence for a cluster population with masses comparable
to giant planets \citep{2000Sci...290..103Z,2000MNRAS.314..858L}. These objects {\it at the bottom 
of the IMF} are inconsistently called planemos, sub-brown dwarfs, cluster 
planets, or isolated planetary mass objects in the literature; in the absence of a definite 
nomenclature, we will use the latter term (abbreviated IPMO) in the following. 

The nature and origin of IPMOs have been debated intensely in recent years. Most of the problems 
raised in the discussions about the formation of brown dwarfs apply even more acutely
to IPMOs \citep[see review by][]{2007prpl.conf..459W}. For example, it is under debate whether 
IPMOs can form directly from the collapse of ultra-low mass cores; instead, they may be giant 
planets ejected from circumstellar disks or stellar embryos ejected from mini-clusters or 
decaying multiple systems. 

As of today, the observational constraints on the nature of IPMOs are very limited. Evidence for 
accretion or disks has been reported for a few young sources with masses around the Deuterium 
burning limit \citep{2002A&A...393..597N,2002ApJ...569L..99Z,2005ApJ...635L..93L,2006ApJ...644..364A,2006ApJ...647L.167J}.
The next step is clearly to establish frequency and lifetimes of IPMO disks, and to push the mass limits
of disk surveys towards the least massive objects in star forming regions. Here we present a Spitzer survey
for disks around IPMOs in the $\sigma$\,Orionis cluster, which harbours the largest IPMO population identified 
thus far. With a well-explored stellar and brown dwarf population \citep{2001ApJ...556..830B,2007A&A...470..903C} 
and a likely age of 2-4\,Myr \citep{2002A&A...384..937Z,2004AJ....128.2316S}, $\sigma$\,Ori is ideal for probing the disk 
frequency as a function of mass as well as the longevity of IPMO disks. 

\section{Observations and photometry}
\label{irac}

As part of the Spitzer program \#30395 (PI: A. Scholz), we obtained deep IRAC images in the four channels centered 
at 3.6, 4.5, 5.8, and 8.0$\,\mu m$ for 18 objects in the $\sigma$\,Ori cluster with estimated masses below 
$20\,M_{\mathrm{Jup}}$. All targets have been identified initially in deep photometric surveys \citep{2000Sci...290..103Z,2002ApJ...578..536Z,2007A&A...470..903C} 
With one exception, all objects have been confirmed as likely ultra-low mass young cluster members with 
low-resolution spectroscopy; however, given the low signal-to-noise of the spectra, this should not be 
taken as a definite membership confirmation. In particular, the membership is under debate for SOri\,70 (see 
Sect. \ref{sori70}) and SOri\,47 \citep{2004ApJ...600.1020M}. See Table \ref{targets_so} for more 
information and references for the target properties. 

The objects, covered with 13 IRAC fields, were observed in a 12 position dither pattern with the medium scale
and 100\,sec integration time per position. In total, this gives 1200\,sec on-source time per target in each
channel, a factor 10 deeper (in signal-to-noise ratio) than the GTO/IRAC survey of this cluster (program \#37, 
PI: G. Fazio, see \citet{2007ApJ...662.1067H}). For the analysis, we made use of the post-BCD mosaics provided by 
the IRAC pipeline version S15.3.0. While parts of some images are affected by saturation effects and straylight from 
bright stars, in the regions where our targets are located the image quality is in most cases excellent.

All objects are detected in channels 1 and 2; two are not detected in channel 3 while four are invisible in channel 4.
For the object SOri\,54, we used the publicly available GTO survey data to measure fluxes in channels 1 and 2,
because in our deep images the source is 'drowned' in the emission from a nearby extended object and
two nearby point-sources (all within 8"). The object SOri\,66 is located on a dark row in our IRAC1 and 2 
images; again we obtained fluxes for this object from the GTO data.

Fluxes were measured by aperture photometry with relatively small apertures (3-5\,pixels) using routines
in {\it daophot}. For all objects, the aperture is free from visible neighbours. Aperture corrections were 
applied, as given in the IRAC data handbook (V3.0). Fluxes were converted to magnitudes 
using the recommmended zeropoints quoted in the data handbook. All IRAC magnitudes for
the IPMOs are reported in Table \ref{targets_so}.

The dominant error source is background uncertainty. To get a realistic estimate, we measured the flux
in empty regions in our images using the same method as for the targets ($>$50 measurements in IRAC1 and 2, 
$>$80 in IRAC3 and 4). As expected, these values scatter around zero; their rms defines the uncertainty in
flux measurements due to background noise and small-scale fluctuations, and thus gives a conservative 
estimate on the photometric error for faint sources. We obtain 1.8, 2.2, 12.7, 20.3$\,\mu$Jy in channels
1-4, respectively. Note that these values do not correspond to the detection limits, which are better
defined using the peak count rate of the PSF in comparison with the background scatter in the sky annulus.
All objects for which fluxes are given in the tables are detected with a signal-to-noise-ratio of at least 4.

The total errors quoted in Tables \ref{targets_so} are comprised of photometric errors 
and systematics. The latter ones are the combination of uncertainties in calibration (5\%) and 
aperture correction (1-2\%, from the IRAC data handbook). No colour correction was carried out; 
instead we added a small contribution (0-2\%) to the error budget. The upper limits quoted in Tables 
\ref{targets_so} correspond to three times the photometric errors. In channels 1 and 2, our fluxes 
are in excellent agreement (within 0.2\,mag) with the values reported recently by \citet{2007A&A...472L...9Z} for
a subsample of our targets. In channels 3 and 4, the magnitude differences between our work and 
\citet{2007A&A...472L...9Z} typically remain within 0.6\,mag, still within the 1$\sigma$ errorbars. We note, 
however, that in these two bands our fluxes are in most cases substantially lower than the values given by 
\citet{2007A&A...472L...9Z}.

\section{Disk Frequency in the Planetary Mass Regime}
\label{diskfreq}

\begin{figure}
\center
\includegraphics[angle=-90,width=10cm]{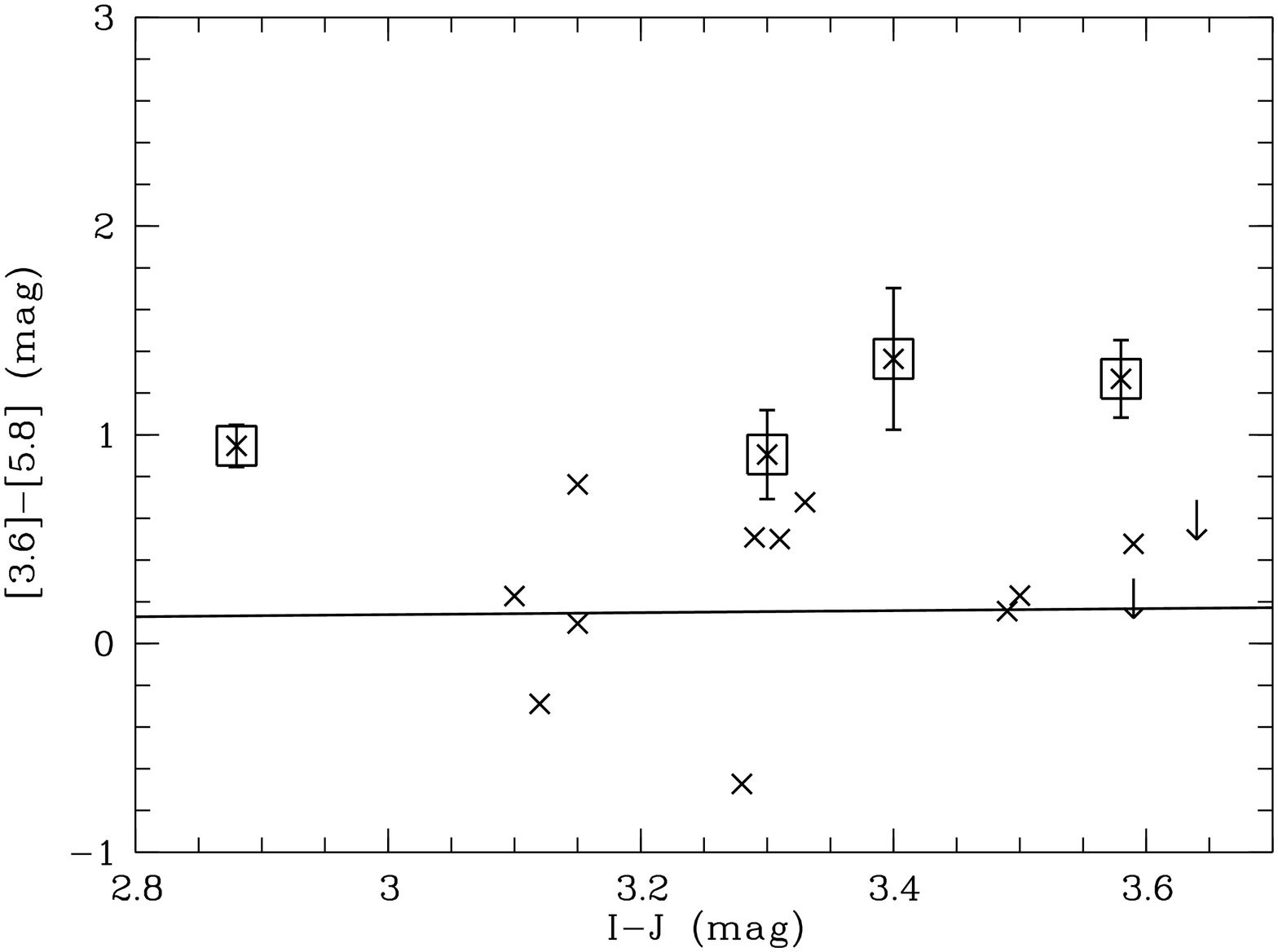}\\
\includegraphics[angle=-90,width=10cm]{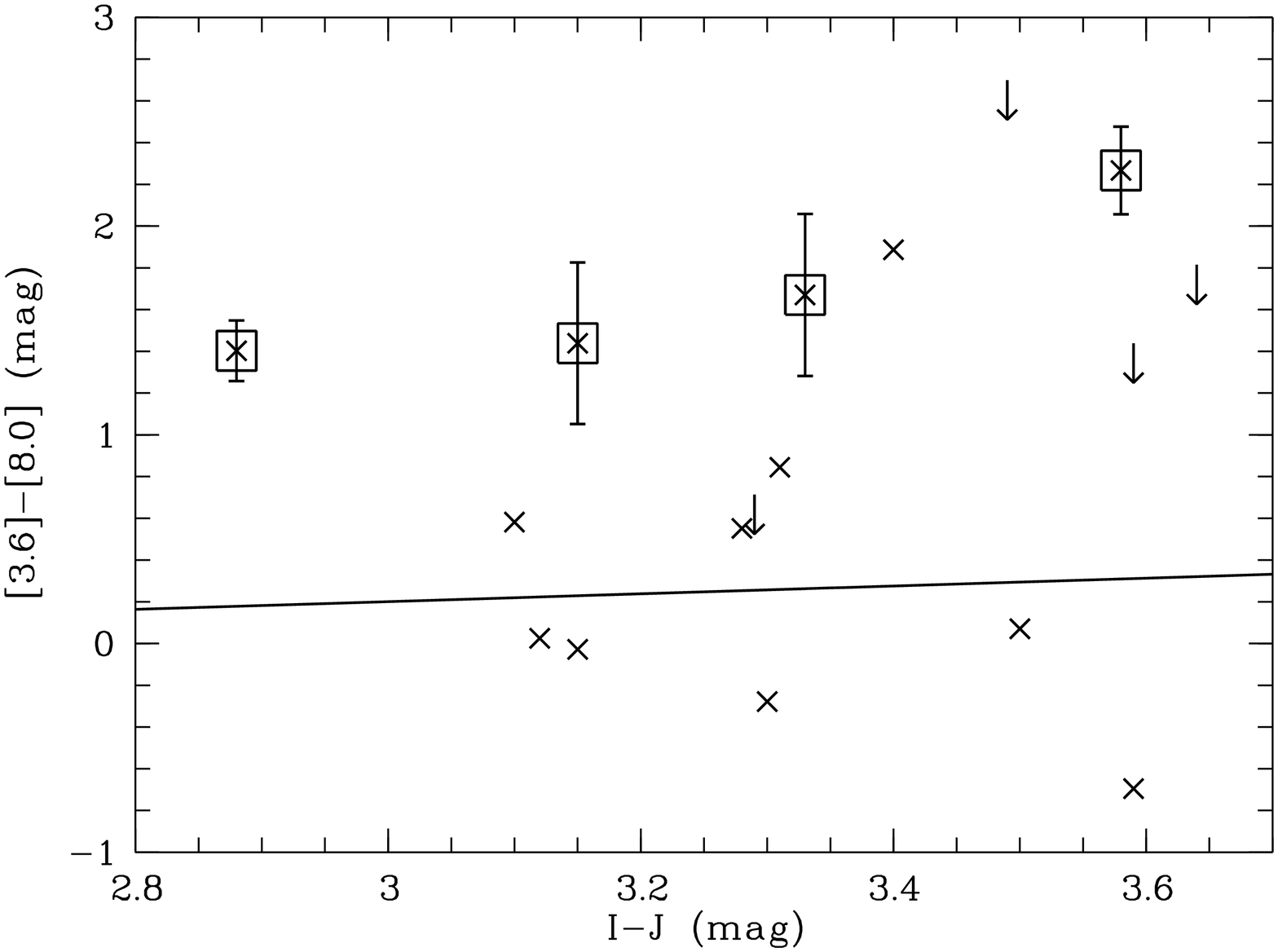}
\caption{IRAC colours for IMPOs in $\sigma$\,Ori (crosses) in comparison with field M/L dwarfs 
\citep[][solid line]{2006ApJ...651..502P}. Objects with $>3\sigma$ excess with respect to the
field dwarfs are marked with squares.
\label{f1}}
\end{figure}

We search for mid-infrared emission from dusty disks in our sample by comparing the measured IRAC 
colours with published values for (disk-less) field objects in the same $T_\mathrm{eff}$ range 
\citep{2006ApJ...651..502P}. The main discriminators to distinguish between the photospheric contribution 
and the disk excess are the fluxes in the IRAC channels 3 and 4; at shorter wavelength the contrast between 
the photosphere and a possible disk is difficult to detect. In Fig. \ref{f1} we plot IRAC colours (I1-I4 and I1-I3) 
for all targets (except SOri\,70, see Sect. \ref{sori70}). As abscissa in these plots we use the $I-J$ 
colour, which can be assumed to be purely photometric and thus represents a proxy for $T_\mathrm{eff}$, 
which in turn correlates with mass for coeval objects. The plot covers a mass range from $\sim 20$ 
to $\sim 8\,M_{\mathrm{Jup}}$. IPMOs are shown as crosses; the solid lines 
are linear fits to the colours of the field objects and thus delineate the photospheric level in these
figures. 

As can be seen in this plot, the majority of the IPMOs have IRAC colours indistinguishable from the
photospheric values and can thus be assumed to have either no disk or only small amounts of 
dust in the inner disk. Sources that lack excess in our data include 
SOri\,54, SOri\,55, SOri\,56 for which \citet{2007A&A...472L...9Z} reported disk detections 
based on the shallow GTO IRAC images. If we use the IRAC3 colour as disk criterion (upper panel), we find 
that four objects -- SOri\,71, SOri\,60, SOri\,66, SOri\,58 -- show excess at the 3$\sigma$ level, i.e. 
4 out of 17 or $24\pm ^{13}_{11}$\%. Since disk excesses may appear only beyond 5.8$\,\mu m$, the result 
from IRAC3 should be considered to be a lower limit. 

Based on the IRAC4 colour (lower panel), we find four sources -- SOri\,71, SOri\,J053949.5-023130, SOri\,60, 
and SOri\,65 -- with $>3\sigma$ colour excesses, which we consider to be primary disk detections.\footnote{The 
object SOri\,66, albeit having a high I1-I4 colour of $1.88$\,mag, is not counted here, because it has a large 
errorbar in IRAC4. A particular case may be the object SOri\,58, which does have excess in IRAC3, but appears
below the photospheric level in IRAC4. This object has been identified to have a disk by \citet{2007A&A...472L...9Z},
based on a IRAC4 flux which is significantly higher than our measurement. Possible reasons for these ambiguous
findings include significant background variations in that area, which might hamper accurate photometry.} 
In all four cases the excess increases significantly in IRAC channel 4 compared with channel 
3, indicating rising flux levels towards longer wavelengths, a clear signature of disk emission. 
With the exception of SOri\,65, these objects have been published previously as disk-bearing very 
low mass sources \citep{2007A&A...470..903C, 2007A&A...472L...9Z}. This gives a disk fraction of 4 
out of 14 or $29\pm ^{16}_{13}$\%. Here we do not count the three objects with upper limits well-above
the photospheric level, for which we cannot decide if they have disk excess or not. Disk frequencies
derived from IRAC3 and IRAC4 are thus consistent; since the value determined from IRAC4 is likely
to be more robust, we put more emphasis on this result. We note that the disk fraction in our sample
might still be somewhat higher than given here, due to the combined effects of photometric uncertainties 
and contaminating field objects \citep[see the dicussion in][]{2007A&A...470..903C}.

We now compare our disk detection rate with previous results for more massive objects in $\sigma$\,Ori, 
as given by \citet{2007ApJ...662.1067H} based on IRAC data: 15\% for HAeBe stars, 27\% for intermediate-mass 
T Tauri stars, 36\% for T Tauri stars, 33\% for brown dwarfs. The value for brown dwarfs is in agreement with 
the disk fraction of 33\% derived by \citet{2003AJ....126.1515J} from ground-based L'-band imaging. A higher 
brown dwarf disk fraction of 47\% has been published by \citet{2007A&A...470..903C}. For the IMPO range 
(objects with $M\lesssim 20\,M_{\mathrm{Jup}}$), we now derive a disk fraction of 29\%, which is compatible 
with the values for T Tauri stars and brown dwarfs within the 1$\sigma$ uncertainties. We do not see a trend to 
higher disk frequencies in the IPMO range, as claimed by \citet{2007A&A...472L...9Z}, instead the evidence points 
to comparable disk fractions for planetary mass objects, brown dwarfs, and T Tauri stars, i.e. over more than 
two orders of magnitude in object mass ($0.008\ldots 2\,M_{\odot}$).

We note that two of the objects with disk excess, SOri\,66 and SOri\,71, stand out from the rest of the
sample, as they show excessively strong H$\alpha$ emission with equivalent widths of $\sim 100$ and
$\sim 700$\AA, respectively \citep{2001A&A...377L...9B,2002A&A...393L..85B}. This indicates that the 
presence of a dusty disks is likely accompanied by ongoing gas accretion, causing intense H$\alpha$ 
emission, as observed in T Tauri stars and brown dwarfs. 

\section{The case of SOri\,70}
\label{sori70}

The faintest target in our sample, SOri\,70, has been reported originally as a $\sim 3\,M_{\mathrm{Jup}}$ 
IPMO with tentative spectral type T5.5 \citep{2002ApJ...578..536Z} . If confirmed, it would be the lowest 
mass free-floating object found to date. Its mass would put it close to or even beyond the predicted opacity 
limit for fragmentation \citep[see][and references therein]{2007prpl.conf..149B}, and thus poses a challenge 
for star formation theory. However, \citet{2004ApJ...604..827B} have questioned the cluster membership (and 
thus the extremely low mass) of SOri\,70, and argue that its near-infrared spectrum is perfectly consistent 
with a foreground dwarf with spectral type $\sim$T6-7. In response, \citet{2004astro.ph.10678M} reinterated 
the claim for cluster membership and youth mainly based on the $(H-K)$ colour of SOri\,70, which is unusually 
high for a T dwarf, and is interpreted as evidence for low surface gravity. To date, the nature of this object
remains unclear. Our deep IRAC images now allow a re-investigation of the issue.

\begin{figure}
\center
\includegraphics[angle=-90,width=10cm]{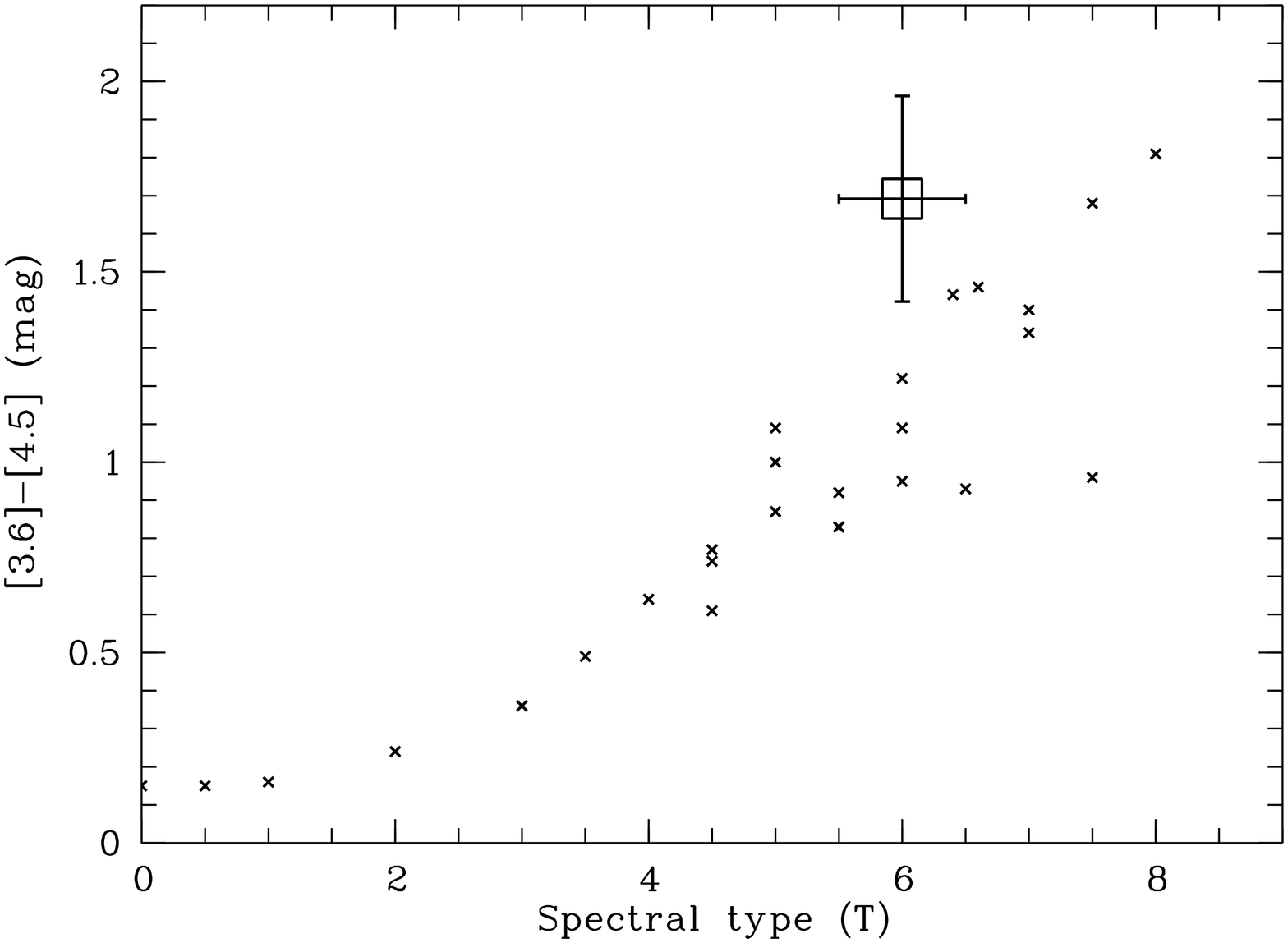}\\
\includegraphics[angle=-90,width=10cm]{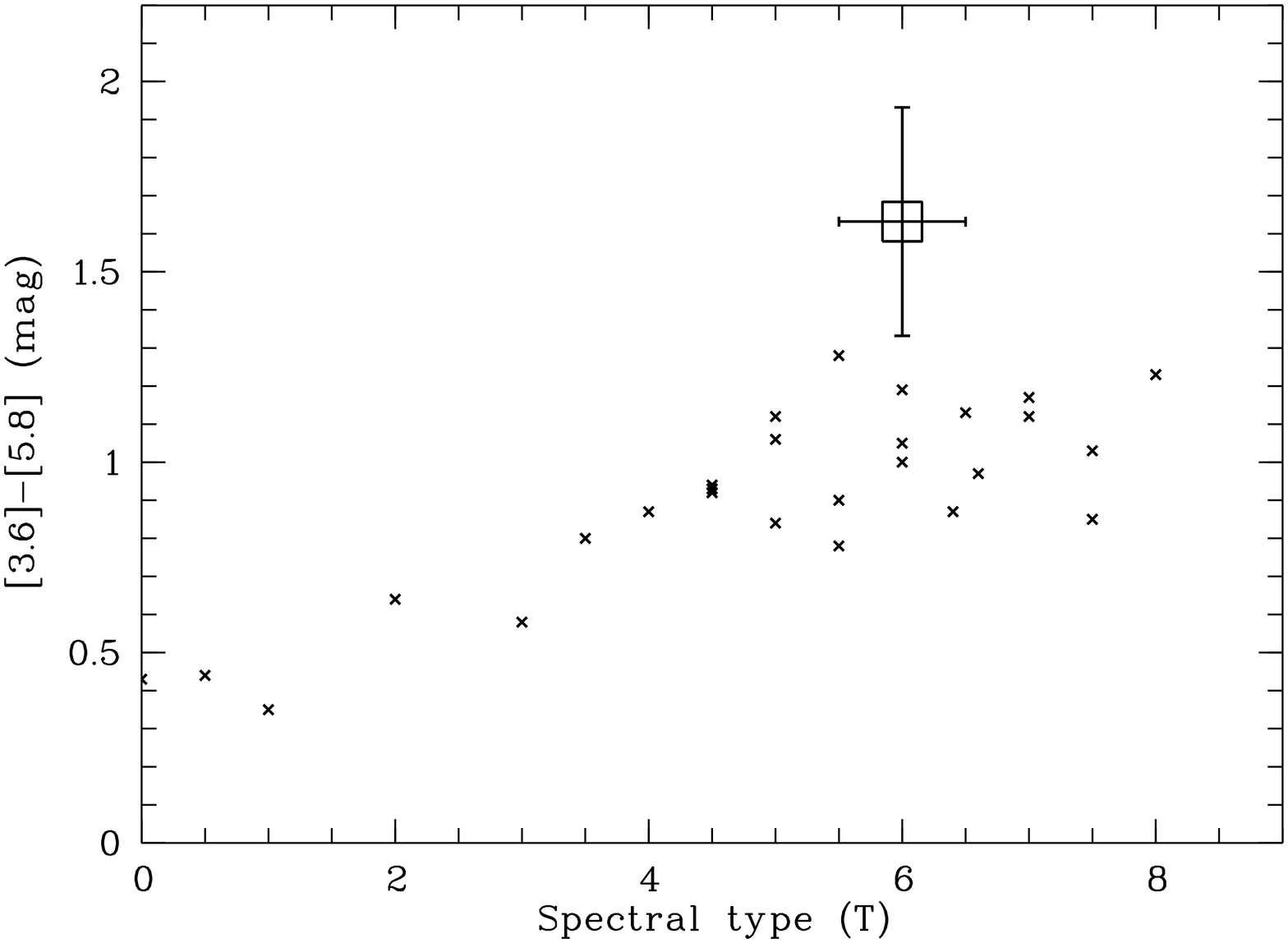}
\caption{The IRAC colours of SOri\,70 (square) compared with field T dwarfs \citep[][small crosses]{2006ApJ...651..502P}. 
\label{f2}}
\end{figure}

SOri\,70 is clearly detected in IRAC channels 1 and 2. In addition, it is detected in IRAC3, 
with a $\sim 4\sigma$ significance (peak countrate over background noise). The IRAC3 flux uncertainty
has been derived based on the very fact that the object is just detected -- if it were $\gtrsim 25$\%
fainter than the estimated 17.2\,mag, we would not be able to see it. In \ref{f2} we plot 
its IRAC colours vs. spectral type in comparison with measurements for field T dwarfs, taken from 
\citet{2006ApJ...651..502P}. SOri\,70 is tentatively plotted spectral type T6 (average of the two
available literature estimates).

As can be seen in this figure, the field T dwarfs form a clear
sequence, whereas SOri\,70 stands out: In both IRAC colours, it appears to have some excess. Compared
with the mean trend for the field T dwarfs, the significance of this excess is $\sim 2\sigma$
in both IRAC2 and IRAC3. This excess cannot be accounted for by uncertainties in the spectral 
type, since the $3.6-5.8\,\mu m$ colour saturates in the late T dwarf regime and reaches maximum values 
of $\sim 1.2$ -- compared with 1.6 for SOri\,70. It should also be pointed out that SOri\,70 is 
brighter than the more massive L5 dwarf SOri\,67 in IRAC2 and 3, again indicating an unusual SED for 
this object.

There are two possible origins for a mid-infrared colour excess in SOri\,70: a) As reported by \citet{2007ApJ...655.1079L},
gravity affects the near/mid-infrared colours of mid/late T dwarfs in the sense that it can produce 
a significant excess at 2-6$\mu m$ for low-gravity objects. b) Similarly to the L type IPMOs in $\sigma$\,Ori,
SOri\,70 might harbour a dusty disk, which produces the IRAC excess.  Both possibilities provide 
additional evidence for youth, and thus the mid-infrared excess bolsters the claim that it is the lowest mass 
free-floating object identified thus far. In case the mid-infrared excess is due to a dusty disk, this can be 
interpreted as evidence for a star-like infancy. While the uncertainties in the IRAC fluxes are too large at this 
stage for a definitive answer, these new findings certainly motivate further follow-up work on this 
target.

\section{Summary}
\label{sum}

We have conducted a disk survey for isolated planetary mass objects, based on deep observations with 
the Spitzer Space Telescope. Our targets comprise the full sample of spectroscopically 
confirmed IPMOs in the $\sigma$\,Orionis cluster. As a pioneering study, the results presented here 
are not intended to be the last word on the subject. Instead, we intend this Letter to serve as an 
important step towards a more profound understanding of the nature and origin of the lowest mass 
objects found in isolation.

A substantial fraction of our targets exhibits mid-infrared excess emission indicative for the existence
of a dusty disk.
From our IRAC data, we derive a disk fraction of $29\pm ^{16}_{13}$\% (4/14) at 8.0\,$\mu m$. A similar
value is found at 5.8\,$\mu m$. These results 
are fully consistent with the disk frequencies derived for T Tauri stars and brown dwarfs in the same cluster. 
The detection of dusty disks around IPMOs in $\sigma$\,Ori is another firm evidence for their nature
as lowest mass members of this young cluster. The finding clearly establishes that objects at the 
bottom of the IMF can harbour dusty disks with lifetimes of at least 2-4\,Myr, the most likely age of the 
$\sigma$\,Ori. Thus, our results fit into previous claims for a T Tauri-like phase in the planetary mass regime 
\citep{2002A&A...393..597N,2005ApJ...635L..93L,2006ApJ...644..364A,2006ApJ...647L.167J}. Disk fractions and 
thus lifetimes are similar for objects spanning more two orders of magnitude in mass (0.008 to 2$\,M_{\odot}$), 
possibly indicating that stars, brown dwarfs, and IPMOs share a common origin. Star formation
theory thus has to account for a number of objects with masses below the Deuterium burning limit.

Our sample includes SOri\,70, the faintest candidate member so far in $\sigma$\,Ori, which is detected
in our IRAC images from 3.6 to 5.8$\,\mu m$. Compared with field T dwarfs, the source stands out based 
on the IRAC colours, with excesses at 4.5 and and 5.8$\,\mu m$ at a $2\sigma$ level. This may be an indication 
for youth and possibly disk occurence for an object with an estimated mass of 3$\,M_\mathrm{Jup}$
\citep{2002ApJ...578..536Z}. Uncertainties are large, though, meaning that this should be seen as motivation 
for further follow-up rather than a definite confirmation of the 'cluster planet' nature of this object.

\acknowledgments	
This work was supported in part by an NSERC grant to RJ.


\clearpage
\newpage

\begin{deluxetable}{llccccccl}
\tabletypesize{\scriptsize}
\tablecaption{Targets and Spitzer photometry in the $\sigma$\,Ori cluster\label{targets_so}}
\tablewidth{0pt}
\tablehead{
\colhead{Name} & \colhead{SpT\tablenotemark{a}} & \colhead{I\tablenotemark{b}} & \colhead{I-J\tablenotemark{b}} & 
\colhead{3.6\,$\mu$} & \colhead{4.5\,$\mu$} & \colhead{5.8\,$\mu$} & \colhead{8.0\,$\mu$} &\colhead{comment} \\
\colhead{} & \colhead{} & \colhead{(mag)} &\colhead{(mag)} & 
\colhead{(mag)} & \colhead{(mag)} & \colhead{(mag)} & \colhead{(mag)} &\\}
\tablecolumns{7}
\startdata
SOri71 & L0        &  20.02  & 2.88 & $15.24\pm 0.06$ &  $14.83\pm 0.06$ & $14.29\pm 0.08$  & $13.84\pm 0.13$ & disk\\	  
SOri47 & L1.5      &  20.53  & 3.15 & $14.87\pm 0.06$ &  $14.91\pm 0.06$ & $14.77\pm 0.11$  & $14.90\pm 0.33$ &\\	  
SOri50 & M9        &  20.66  & 3.12 & $15.66\pm 0.06$ &  $15.64\pm 0.06$ & $15.95\pm 0.30$  & $15.63\pm 0.70$ &\\	  
SOri51 & M9        &  20.71  & 3.50 & $15.42\pm 0.06$ &  $15.30\pm 0.06$ & $15.19\pm 0.15$  & $15.35\pm 0.51$ &\\	  
SOri53 & M9        &  21.17  & 3.28 & $16.02\pm 0.06$ &  $15.84\pm 0.06$ & $16.69\pm 0.63$  & $15.47\pm 0.58$ &\\	  
SOri54 & M9.5      &  21.29  & 3.30 & $15.8\pm 0.1$\tablenotemark{c} &  $\sim 15.6$\tablenotemark{c,d} & $\sim 15.3$\tablenotemark{d}  & $>15.1$ \\   
SOri55 & M9        &  21.32  & 3.10 & $16.41\pm 0.06$ &  $16.19\pm 0.07$ & $16.18\pm 0.37$  & $15.83\pm 0.90$ &\\	  
SOri56 & L0.5      &  21.74  & 3.30 & $16.12\pm 0.06$ &  $15.79\pm 0.06$ & $15.62\pm 0.22$  & $15.27\pm 0.48$ &\\	  
SOri58 & L0        &  21.90  & 3.30 & $16.43\pm 0.06$ &  $16.19\pm 0.07$ & $15.52\pm 0.20$  & $\sim 16.7$\tablenotemark{d} & disk?\\	  
SOri\,J053949.5-023130 &NA &  22.04  & 3.15 & $16.49\pm 0.06$ &  $16.32\pm 0.07$ & $15.73\pm 0.25$  & $15.05\pm 0.38$ & disk\\
SOri60 & L2        &  22.75  & 3.58 & $16.61\pm 0.06$ &  $16.34\pm 0.07$ & $15.34\pm 0.18$  & $14.34\pm 0.20$ & disk\\	  
SOri62 & L2        &  23.03  & 3.59 & $16.50\pm 0.06$ &  $16.36\pm 0.07$ & $16.02\pm 0.32$  & $\sim 17.2$\tablenotemark{d} &\\	  
SOri65 & L3.5      &  23.23  & 3.33 & $16.72\pm 0.07$ &  $16.60\pm 0.08$ & $16.05\pm 0.33$  & $15.05\pm 0.38$ & disk\\	  
SOri66 & L3.5      &  23.23  & 3.40 & $17.41\pm 0.1$\tablenotemark{c}  &  $16.87\pm 0.1$\tablenotemark{c} & $16.05\pm 0.33$  & $15.53\pm 0.62$ & disk?\\ 
SOri67 & L5        &  23.40  & 3.49 & $17.75\pm 0.10$ &  $17.64\pm 0.16$ & $\sim 17.6\tablenotemark{d}	$   & $>15.1$ &\\	  
SOri68 & L5        &  23.77  & 3.59 & $16.49\pm 0.06$ &  $16.39\pm 0.07$ & $>16.2$	    & $>15.1$ &	     \\	  
SOri69 & T0        &  23.89  & 3.64 & $16.87\pm 0.07$ &  $17.14\pm 0.12$ & $>16.2$	    & $>15.1$ &	     \\	  
SOri70 & T         &  25.03  & 4.75 & $18.83\pm 0.25$ &  $17.14\pm 0.11$ & $\sim 17.2$\tablenotemark{d}      & $>15.1$ &    \\	  
\enddata
\tablenotetext{a}{spectral types from \citet{2001ApJ...558L.117M,2001A&A...377L...9B,2002ApJ...578..536Z,2002A&A...393L..85B}}
\tablenotetext{b}{optica/NIR photometry from \citet{2000Sci...290..103Z,2002ApJ...578..536Z,2007A&A...470..903C}}
\tablenotetext{c}{fluxes measured in GTO images}
\tablenotetext{d}{uncertainty is approximately $\pm 0.5$\,mag}
\end{deluxetable}

\clearpage

\clearpage
\newpage

\end{document}